# Silicate dielectric ceramics for millimetre wave applications

Franz Kamutzki[1*], Sven Schneider[1], Jan Barowski[2], Aleksander Gurlo[1], Dorian A. H. Hanaor[1]

[1] *Fachgebiet Keramische Werkstoffe, Technische Universität Berlin, 10623 Berlin, Germany*

[2] *Lehrstuhl Hochfrequenzsysteme, Ruhr-Universität Bochum, 44801 Bochum, Germany*

*\* Corresponding author: franz.kamutzki@ceramics.tu-berlin.de*

## Abstract

Silicate ceramics are of considerable promise as high frequency dielectrics in emerging millimetre wave applications including high bandwidth wireless communication and sensing. In this review, we show how high quality factors and low, thermally stable permittivities arise in ordered silicate structures. On the basis of a large number of existing studies, the dielectric performance of silicate ceramics is comprehensively summarised and presented, showing how microstructure and $SiO_4$ tetrahedral connectivity affect polarizability and dielectric losses. We critically examine the appropriateness of silicate materials in future applications as effective millimetre wave dielectrics with low losses and tuneable permittivities. The development of new soft chemistry based processing routes for silicate dielectric ceramics is identified as being instrumental towards the reduction of processing temperatures, thus enabling silicate ceramics to be co-fired in the production of components functioning in the mm wave regime.

## 1. Introduction

Ceramic dielectric materials have underpinned the development of wireless technologies in recent decades. With the rapid extension of internet connectivity to an increasingly wide variety of electronic devices (Internet of things, IoT) and the introduction of new high bandwidth network technologies such as 5G telecommunication, intelligent transport systems (ITS), remote health care and ultra-high-definition television (UHDTV), wireless data traffic is expected to explode ~10000 fold over the coming years [1]. To accommodate this massive increase, wireless data transmission in the millimetre wave (mmW) range from 30 – 300 GHz and even higher will have to be utilized, with an associated shift in dielectric materials employed in various components. Relative to frequencies in the regime 5-30 GHz, mmW frequencies offer exponentially higher bandwidth at lower component and narrower beam sizes [2]. To comprehend the remarkably rapid increase in mobile data traffic driven by device connectivity and wireless broadband innovations, a few numbers can be helpful:

- Having reached about 15.4 billion in 2015, the number of worldwide installed IoT devices is expected to exceed 75 billion in 2025 [3]
- IoT Technologies could generate up to 14.4 trillion U.S. $ in value [4] and the total investment in this market is predicted to reach 15 trillion $ by 2025 [5]
- In 2013 the global mobile data traffic volume was projected to reach 11.2 exabyte (EB) per month by 2017 [5]. This estimation was surpassed, with 11.51 EB per month recorded in that year [6].
- The latest figures of the International Telecommunication Union (ITU) predict an annual growth rate of at least 55% in wireless data transfer volume, with the uptake of 5G technologies, connected

- device communication as well as driver assistance and autonomous vehicle transportation [7].
- This projected growth rate would yield an annual traffic of 368 000 EB or $3.68 \times 10^{23}$ bytes in 2035 (Figure 1). This estimation can be considered conservative given that an introduction of prospective 6G technology with an increased data transfer rate of x100 compared to 5G is already on the horizon [7].

Components used in contemporary mmW applications are mostly based on materials developed originally for use in lower frequency microwave systems. Rapid progression in the development of mmW based communication and radar systems for autonomous vehicles drives the need for novel ceramic dielectric components in the form of dielectric resonators, filters, antennas, waveguides and substrates that are better suited to effective high frequency implementation and wide-scale production. In particular, materials with improved dielectric loss characteristics will be in great demand, as currently employed materials in lower frequencies will not be implementable in the high GHz range where dielectric losses will be enhanced. Besides appropriate dielectric properties, materials for such applications should exhibit low sintering temperatures to facilitate their implementation in the form of low temperature co-fired ceramics (LTCCs), which are often applied for the fabrication of integrated components and substrates that allow simultaneous processing of multiple materials.

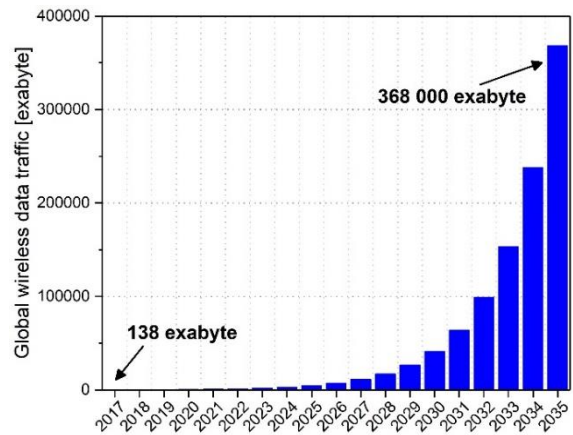

*Figure 1: Expected annual global wireless data traffic based on a forecast by the ITU [7].*

Dielectrics tend to be categorised on the basis of their relative permittivity $\varepsilon_r$, also referred to as the dielectric constant $k$, which describes a material's absolute permittivity $\varepsilon$ relation to the vacuum permittivity $\varepsilon_0$

$$k = \varepsilon_r = \frac{\varepsilon}{\varepsilon_0} \qquad (1)$$

Here these terms are used interchangeably. Low loss dielectric oxide ceramics with intermediate values of $\varepsilon_r$ have long been of special interest for devices utilising microwave frequencies encompassing applications in cellular phones, satellite communication, and wireless LAN. In dielectric resonators operating at frequencies up to 10 GHz, low loss perovskite materials with $\varepsilon_r$ values in the range 20-100 are commonly used to enable the functionality and miniaturisation of components. However, for dielectric components operating in the mmW range and beyond, material requirements differ somewhat. At mmW frequencies, wavelengths and circuit dimensions shrink and more stringent tolerances are required in circuit material parameters. In particular, with a shift to higher frequencies (> 30GHz), resonator miniaturization is less of a concern and lower permittivity materials are desirable with the minimization of dielectric losses and stability of the effective $k$ values being of even greater importance. Well-controlled permittivities and low levels of dielectric losses are of accentuated importance also in passive mmW components such as nonradiative waveguides and substrates.

Because of their crystal chemistry, a special class of oxides, namely the silicates, naturally possess low $\varepsilon_r$, potentially low levels of dielectric loss and are relatively resistant to temperature changes both in performance and mechanical properties. Moreover, many known silicate materials, both natural and synthetic ones, consist of abundant alkaline, alkaline earth metals and silicon, making their production inexpensive and reliable. Consequently, silicates are increasingly the subject of materials research towards novel high frequency applications. The interested reader is able to find multiple reviews summarizing in various lengths microwave and millimetre wave ceramic materials, including their production, characterization and applications [8–20]. However, the development and application of silicate ceramics for extremely high frequency applications is still in early stages and merits closer examination.

## 2. Millimetre wave technologies

The mmW region of the electromagnetic spectrum underpins many emerging technologies and is designated by the ITU as the Extremely High Frequency, or EHF Band, spanning the region from 30 to 300 GHz, as illustrated in Figure 2. The interest in millimetre wave technologies and the consequent demand for cheaply and sustainably producible dielectric materials that meet stringent performance requirements, is driven by significant shifts in our contemporary technological environment. Recent decades have been marked by highly consequential advancements in device connectivity, uptake of unmanned aerial vehicles, mobile communication devices, high volume wireless data transfer and autonomous vehicles. Enabled by advances in computer vision and AI, these technologies are expected to continue to grow rapidly in the foreseeable future, and require high resolution sensing capabilities and large bandwidth transmission in order to achieve their full potential.

To meet the requirement for high multi-directional spatial resolution in adverse weather conditions and rapid data transfer between large numbers of devices, vehicles and sensors, it is a logical approach to utilise high frequency microwaves, namely those in the mm wave portion of the spectrum, as indicated by Figure 2.

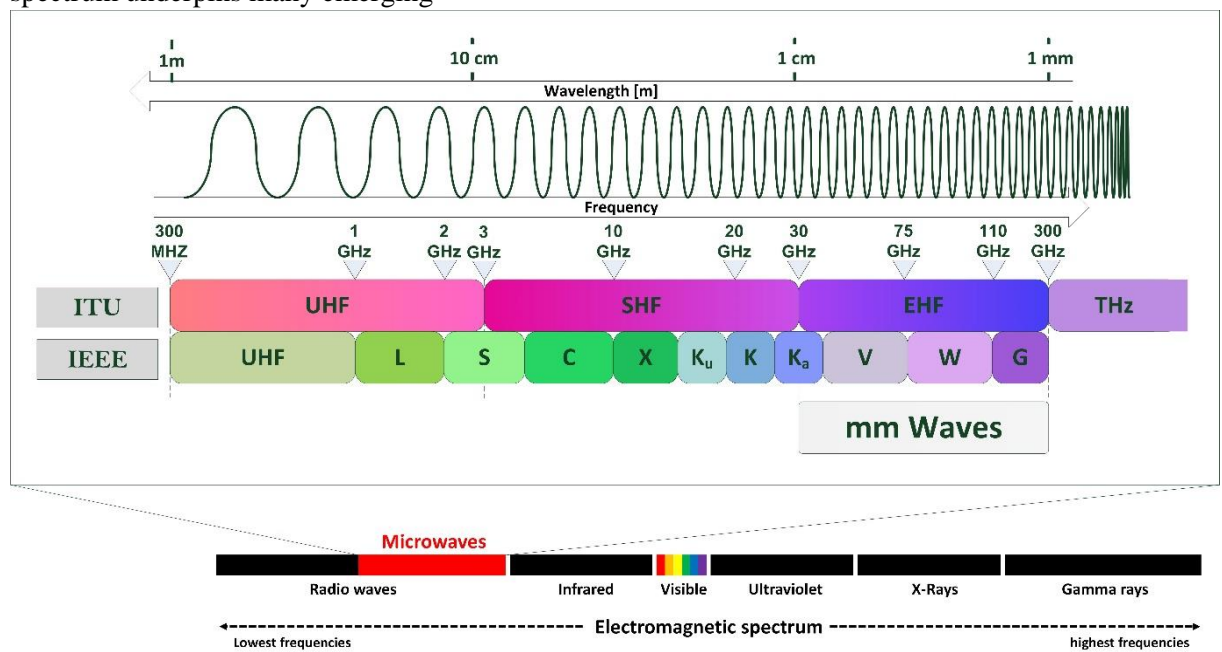

*Figure 2: The microwave portion of the electromagnetic spectrum and its constituent bands [21].*

The shift towards higher frequencies in the mm wave region stems from advantages in terms of bandwidth and resolution. Owing to its small wavelengths, the uptake of mm wave radar in the 1990s enabled ground and air based systems to detect smaller objects with higher precision than with earlier radar systems designed for detecting large highly reflective objects. The role of Radar sensors in the future automotive industry cannot be overestimated and is identified as part of the so-called 'Automotive Megatrend' of autonomous driving and safety [22] . Existing vehicular millimetre-wave radar applications include assisted parking, remote parking, auto pilots, traffic jam pilots, lane change assist, blind spot detection, pre-crash rear Radar, rear traffic alert, front-cross-traffic-alert and exit assist function. As vehicles become increasingly autonomous the importance and number of high frequency radar systems, currently planned at frequencies of 80 and 140 GHz, is set to grow.

In wireless communication, the high frequencies of mm waves provide the bandwidth necessary for the transfer of large amounts of data. Such mm waves enable fifth generation communication (colloquially 5G) and are generally transmitted in narrow beams over comparatively short distances relative to longer microwave transmissions. Disadvantages of communication through mm waves lie in their short transmission length and severe signal attenuation, with waves easily blocked by obstacles, resulting in small coverage. The applications of mmW and other microwave frequencies are schematically illustrated in Figure 3.

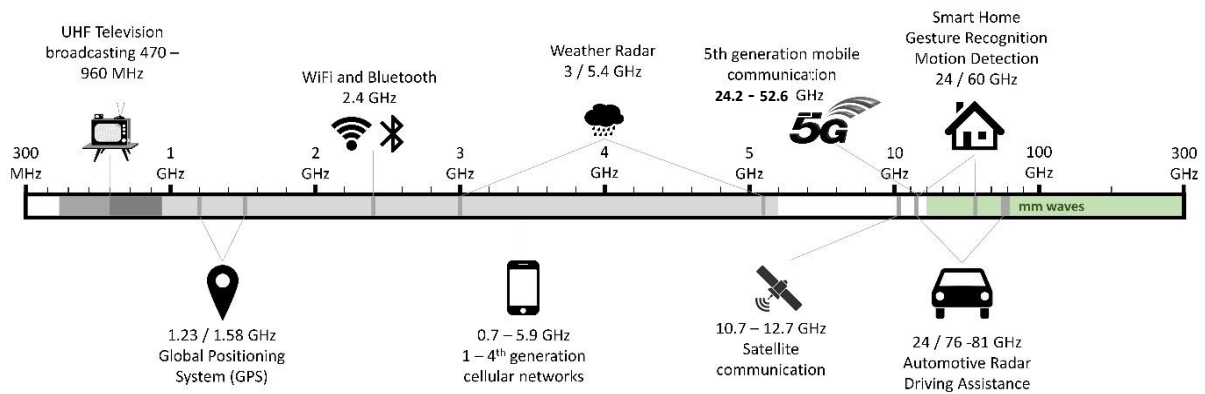

*Figure 3: Applications using microwave frequencies. Data collected from [23–26].*

The short range and narrow beam length of mm wave frequencies further contribute to the need for large numbers of transmitting and receiving elements in the uptake of new communication protocols, and further motivates the search for alternative high performance mm wave dielectric ceramics.

Towards the effective utilisation of higher frequency microwaves, mass production of well-performing millimetre wave dielectrics functioning in the range of frequencies from 30 to 300 GHz is of great importance. Table 1 provides an overview of typical components in millimetre wave technologies. A taxonomy of mmW communication related research and applications is given by Wang et al. [2], in which they further distinguish between different mmW applications in (i) wearable devices, (ii) virtual reality, (iii) vehicular networks, (iv) satellite communication, (v) 5G communication, (vi) object imaging and tracking and (vii) object detection.

One substantial obstacle to the forecasted paradigm shifts in transport and communication is the foreseeable inability of industry to supply high performance sensor substrates, antennae, resonators and high power energy storage devices at costs and scales that allow their widescale uptake without compromising attainable performance. Currently, electronic components account for a significant portion of vehicle production costs, with this fraction set to rise due to supply chain inelasticity [27,28]. As an example, the Tesla Model 5 has approximately 10.000 installed multi-layer ceramic capacitors (MLCCs), which is around 5 times more than in a car of comparable class with combustion engine [29]. Similarly, the number of MLCCs doubled from the iPhone 6s to the iPhone X from 500 to 1000. The demand for MLCCs has outstripped the supply for many years now, leading to increasing lead times and prices for many consumer electronics. The compositional diversity of silicate ceramics offers the possibility to tailor the dielectric properties of material systems to individual applications [30]. In an industrial context, dielectric antennae [31] or antenna-lens combinations [32] offer the possibility to shape radiated beam profiles according to system needs. Furthermore, they can be used to realize chemical or mechanical decoupling between the electronics and the monitored process. For this purpose, geometry and dielectric properties must be well controlled to ensure correct refraction properties.

*Table 1: Typical components in millimetre wave technologies.*

| Millimeterwave components & devices | |
|---|---|
| **Passive** | **Active** |
| Resistors*, Inductors, Capacitors*, Filters*, Antennas*, Packages*, Waveguides*, Circuit Board Substrates*, Diodes, Resonators* | Varactors, Frequency multipliers, Transistors, Amplifiers, Phase- & frequency discriminators, Isolators, Signal synthesizers, Mixers |

\* denotes a device that can be realized using ceramics

## 3. Requirements of millimetre wave dielectrics

Three key parameters are of primary interest in the evaluation of dielectric materials for millimetre wave frequencies. These are the relative permittivity $\varepsilon_r$, quality factor $Q_f$ and temperature coefficient of resonant frequency $\tau_f$. In general, a higher $\varepsilon_r$ value is desired in applications where component miniaturisation is critical, e.g. in the high MHz and low GHz range, e.g. for cell phones, as high permittivity allows for smaller component dimensions according to Equation 3. However, for the millimetre wave region of the spectrum miniaturisation is not critical anymore and low values of $\varepsilon_r$ are sought after because the low polarizability of such materials under a high frequency field reduces noise and inductive cross-talk and further allows for improved signal propagation velocity inside the material, a term given by:

$$v_p = \frac{c}{\sqrt{\varepsilon_r}} \quad (1)$$

where $v_p$ is the propagation velocity (m/s) and c is the speed of light (m/s). The target range for dielectric constants in mmW components depends on the application and operating frequency. This dependence can be illustrated using the example of a cylindrical dielectric resonator, an essential component in microwave technologies. The concept of a dielectric resonator (DR) was established by Richtmyer in 1939 [33] by his discovery that dielectric, non-metallic materials can function as resonators in the microwave range. A typical DR consists of a ceramic block, having one of various standard shapes mounted on a substrate (Figure 4a). Radiowaves can be introduced into the ceramic via different techniques such as a coaxial or microstrip feed and can be confined within the material, due to the permittivity change at the ceramic-air interface, forming a standing wave at resonant frequency. However, because the DR walls are partially transparent for radio waves, the resonating waves can also be radiated into free space [34], which may introduce undesired signal losses, if the resonator is not properly designed.

The attributes of an ideal DR include small dimensions, very low dielectric loss, low temperature coefficient of resonant frequency ($\tau_f$) and compatibility with integrated circuits, which render them significantly more attractive compared to earlier systems of metallic cavity resonators. In particular, DRs are a logical choice for extremely high frequency applications, where metallic cavities become too lossy. Ceramic dielectrics are most commonly employed as either dielectric resonator antennae (DRAs) or dielectric resonator oscillators (DROs). Initially, DRAs were applied predominantly in rectangular shapes, before Long et al. proposed the cylindrical dielectric cavity antenna, which proved to be advantageous in the mmW region [35]. Many novel DRA applications have been introduced in recent years, offering great improvements in terms of bandwidth, gain and coupling [36]. Depending on the intended antenna application, many different geometries, parameters, excitation methods and resonant modes are employed in DRAs. Recently Mukherjee et al. [37] summarized in detail the advancements in this field. A DRO is an oscillator using a dielectric resonator material to transmit signals with very high stability. Similar to DRAs, effective ceramic DROs require attributes of low dielectric loss, excellent temperature stability, small size and usually low cost, making them exceptionally suitable for millimetre wave frequencies. Further, the progress in the design and implementation of dielectric resonators as filters has been summarized recently by Dubey et al. [38].

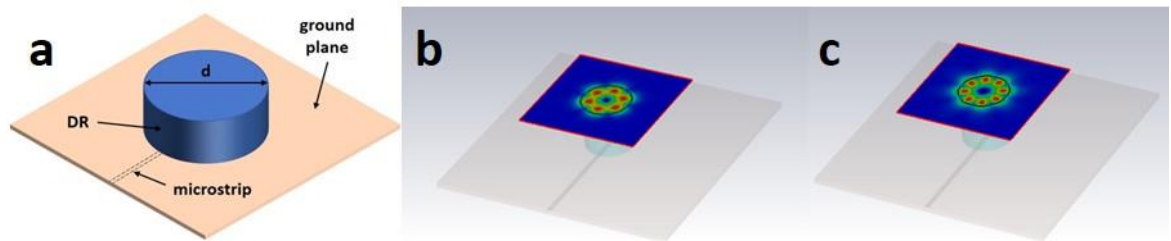

*Figure 4: A schematic illustration (a) of a ceramic dielectric resonator fed by a microstrip line on a printed circuit board substrate. The incident field excites so called "whispering gallery modes" (b) and (c), which are eigenmodes of the dielectric cylinder. By increasing the dielectric constant at constant dimensions from (b) to (c) higher order modes are excited.*

The resonant frequency of a dielectric resonator in free space is a function of its dimensions, shape and resonance mode and the material's permittivity, but can be simplified according to the general concept

$$f_0 = \frac{c}{d\sqrt{\varepsilon_r}} \quad (2)$$

Where $f_0$ is the resonant frequency and $d$ the diameter of the dielectric resonator. From Equation 2 and 3 one can see that lower permittivities yield higher propagation speeds and high resonant frequencies. For the specific case of a cylindrical resonator such as that depicted in Figure 4a, the relationship between resonator dimensions, $\varepsilon_r$ and the frequency of the dominant resonant mode is given by equation 4 in [39], with the Bessel functions tabulated in Keyrouz et al. [40], and is visualized in Figure 5, showing the relationship between resonator size, frequency and permittivity. A low permittivity material can be realised at a macroscopic scale by a decreased density (increased porosity) or at an atomic scale by decreased polarizability [41]. For the case of high frequency dielectrics with a set operating frequency and constrained component dimensions, one can see from Eq. 3 that it is $\varepsilon_r$ that determines the resonant frequency. It is therefore useful to have a predictive theory about the permittivity of an unknown material composition. The permittivity of a material is related to its dielectric polarizability according to the Clausius-Mosotti equation (CME) [42]

$$\alpha_D^T = \left(\frac{1}{\frac{4\pi}{3}}\right)\left[\frac{(V_m)(\varepsilon_r - 1)}{(\varepsilon_r + 2)}\right] \quad (3)$$

where $\alpha_D^T$ is the dielectric polarizability including both ionic and electronic components, and $V_m$ the molar volume of the compound ($\text{Å}^3$). Roberts demonstrated the applicability of the Clausius Mosotti relation for the estimation of dielectric constants for unknown compounds [43]:

$$\varepsilon_C = (3V_m + 8\pi\alpha_D^T)/(3V_m - 4\pi\alpha_D^T) \quad (5)$$

With $\varepsilon_C$ as calculated dielectric constant and $\alpha_D^T$ the total dielectric polarizability of the compound. The value of $\alpha_D^T$ for oxide materials can be calculated using a set of ion polarizabilities developed by Shannon [44].

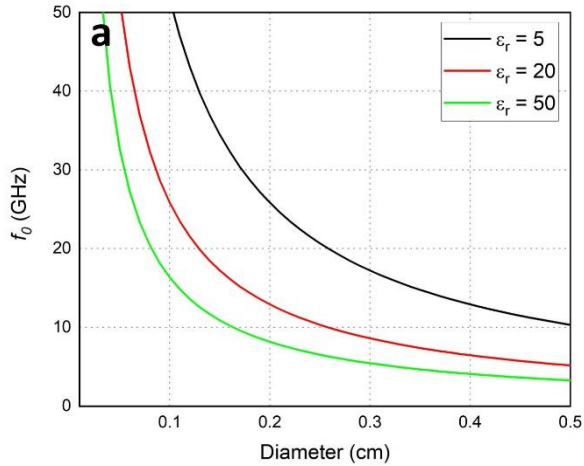
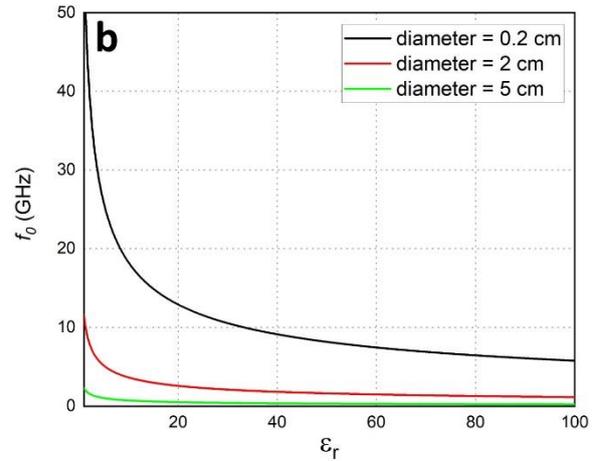

*Figure 5: Resonant frequencies of a cylindrical resonator as a function of (a) the diameter of the resonator for three different $\varepsilon_r$ values and (b) as a function of the resonator materials' $\varepsilon_r$ for three different geometries.*

Another approach for the prediction of $\varepsilon_r$ was developed on the basis of the P-V-L model by Philips, van Vechten and Levine for the evaluation of bond covalencies and ionicities in binary crystals [45–50]. This model was generalized by Wu et al. by decomposing complex crystals into the sum of their constituent binary crystals [51]

$$A_{a1}^1 \cdot A_{a2}^2 \ldots A_{ai}^i \ldots B_{b1}^1 \cdot B_{b2}^2 \ldots B_{bj}^j = \sum_{i,j} A_{mi}^i \cdot B_{nj}^j \quad (6)$$

In which the complex crystal is represented by the left side of the equation with cations $A$ of element $i$ with $ai$ the corresponding elements and anions $B$ if element $j$ with corresponding elements $bj$. The right side represents the decomposed binary crystals with cations $A$ of element $i$ with the coordination share $mi$ and anions $B$ of element $j$ with coordination share $nj$. For every binary crystal, the atomic bonds μ can be quantized in ratios of covalent and ionic character ($f_c^\mu, f_i^\mu$) which both contribute to the average energy gap $E_g^\mu$ of the bond μ. Eventually, the dielectric character of each bond can be described by means of its susceptibility

$$\chi_b^\mu = \frac{1}{4\pi}\frac{(h\Omega_p^\mu)^2}{(2\pi E_g^\mu)^2} \quad (7)$$

with $\Omega_p^\mu$ as the plasma frequency of bond μ. The macroscopic linear susceptibility $\chi$ can be calculated from all $\chi_b^\mu$ and the number of μ-type bond per cubic centimeter ($N_b^\mu$)

$$\chi = \sum_\mu N_b^\mu \chi_b^\mu \quad (8)$$

The relative permittivity is then $\varepsilon_r = \chi + 1$.

Using either the Clausius and Mosotti or the P-V-L methods, the tunability of $\varepsilon_r$ by ion substitution in a silicate lattice or other material can be assessed theoretically. In contrast to the CME, by considering the local structural environment of individual cation and anion sites, the more complex P-V-L based approach accounts for anisotropy in the lattice structure and should yield greater accuracy in the prediction of bulk dielectric polarizability. A comprehensive examination of the validity of the CME based method for the prediction of dielectric constants in oxides was conducted by Shannon. This study revealed that many simple compounds, including some silicates, can be modelled well simply on the basis of ionic polarizabilities and molar volumes. However, this study also found large discrepancies of experimental behaviour from predicted behaviour for silicates from the sorosilicate, tectosilicate and pyroxene families [44]. The ability to predict dielectric performance in silicate systems is key to the design of new mmW silicate dielectric ceramics.

A second, more stringent requirement of mmW dielectric materials is a low dielectric loss corresponding to a high ratio of energy storage to loss. The resonator in Figure 4a is excited by a microstrip transmission line, which comprises a smaller conductor on the top layer and a full ground plane on the opposite side of the substrate. In this case, certain eigenmodes of the dielectric cylinder "whispering gallery" modes, as shown in Figure 4 (b) and (c), are aimed at. The performance of the resonator, namely its *quality factor*, is therefore directly linked to the material losses. The total dielectric loss is the sum of intrinsic and extrinsic loss mechanisms. The term quality factor, commonly denoted by $Q$ is often used as a measure of the power loss in microwave resonators and is defined as the reciprocal of the loss tangent $\tan \delta$:

$$Q = \frac{1}{\tan\delta} \qquad (9)$$

where $\delta$ is the loss angle. In practice, a smaller loss angle leading to higher $Q$ values means less dissipated energy in the dielectric material. Usually, there is a linear relationship between dielectric losses and the frequency, making the introduction of $Q_f$ useful:

$$Q_f = \frac{f_0}{\tan\delta} \qquad (10)$$

In dense ceramics at lower frequencies $Q_f$ is effectively constant However, at higher frequencies in the microwave and millimetre range it increases somewhat with frequency [52]. It should be noted here that the frequency dependence of $Q_f$ for low loss, high frequency dielectric materials is a lot smaller than that for giant permittivity materials. Sebastian et al. also pointed out, that small samples resonating at higher frequencies statistically have fewer defects, simply because of their smaller geometries [15]. Nonetheless, it is suggested that the resonant frequency should always be noted when reporting $Q_f$ values. By proper ceramic processing, extrinsic losses can be reduced to a minimum in ceramic dielectrics. However, because of the complex nature of extrinsic and intrinsic dielectric loss mechanisms and its frequency dependence, it is very challenging to predict $Q$ especially at high frequencies. The optimisation of $Q_f$ in the field of microwave and millimetre wave materials is therefore still very experimental with a trial and error character. An overview of different materials research areas based on $\varepsilon_r$ and attained $Q_f$ values was given by Ohsato et al. (Figure 6) and cleary shows that ultra high $Q_f$ are only realisable for very low permittivity materials [53].

A third requirement that is often overlooked is a near-zero temperature coefficient of resonant frequency $\tau_f$, corresponding to the sensitivity of the resonant frequency to temperature. Changes in the effective permittivity of a dielectric applied in a circuit, due to environmental parameters, including thermal fluctuations and water adsorption, cause variations in impedance, which tend to be more significant at higher frequencies. For practical implementation of a high frequency material, its resonant frequency should be independent of a change in temperature to ensure proper functioning in variable conditions. $\tau_f$ is a composite parameter

$$\tau_f = -(\frac{1}{2}\tau_\varepsilon + \alpha_L) \qquad (11)$$

related to the materials drift of permittivity with temperature $\tau_\varepsilon$ and thermal expansion $\alpha_L$. As thermal expansion values in electronic ceramics are generally small, $\tau_\varepsilon$ is the determining factor for $\tau_f$. It follows from Equation 11 that if $\tau_f$ should be $\approx 0$ it is then required that $\tau_\varepsilon \approx 2\alpha_L$. Adjusting a material's $\tau_f$ towards zero can be achieved either chemically, through doping or substituting, or by forming composite materials with opposing $\tau_f$ values. The latter has been demonstrated manifold in the literature [54–58], e.g. by mixing materials with negative $\tau_f$ with a small weight fraction of very large $\tau_f$-materials like $TiO_2$. This method is simple and predictable, however the products often suffer from low Q values due to the introduction of defects into the microstructure. A similar method was proposed by Sebastian et al. by stacking resonators of different materials with opposing $\tau_f$ values directly on top of each other [59]. Tuning a material's $\tau_f$ within a single phase should be preferred since it reduces interface polarization and allows for minimization of dielectric losses. However, this is only possible via doping (only a minimal amount of foreign atoms in the lattice) or when the mixed endmember materials form a solid solution. Interestingly, a recent study by Zou et al. found that sintering $Ba_2ZnSi_2O_7$ in an atmosphere of $N_2 + 1$ vol% $H_2$ could bring about an improvement in $\tau_f$ values. [60]

## 4. Characterisation of dielectric properties in the mmW region

Measuring the dielectric properties of materials in the millimetre-wave range from 30 GHz to 300 GHz and sub-THz Wave range above 300 GHz can be challenging, as the comparably short wavelength (below 10mm) of such frequencies entails sensitivity to mechanical or geometrical misalignments which may quickly dominate the measurement result.

While standard impedance spectroscopy extends to the RF range, other methods are needed for mm wave parameters. In general, one may differentiate between resonant and broadband measurement methods, as well as guided-wave or free-space methods. Depending on the demands for specimen preparation, required accuracy, and frequency range, a suitable combination of these parameters has to be identified. An extensive overview of the existing methods is given in [61].

Resonant methods utilise a highly accurately manufactured specimen in a resonant structure, usually a cavity. From the change in resonance frequency due to the presence of the specimen, the relative permittivity of the sample can be identified. Additionally, a highly precise characterization of the electromagnetic losses within the sample can be done by evaluating the resonators quality factor, i.e. the depth of the resonance. The measurement is usually done with a vector network analyser that performs a stepped frequency sweep over the frequency range of interest. In case of the Hakki Coleman (Courtney) dielectric resonator method [62], a Cylindrical (puck) specimen is placed between copper plates and the resonators' resonant behaviour can be measured either in transmission or reflection.

Methods that provide greater freedom in sample shape are usually free-space or transmission-line methods. These examine the reflected and/or the transmitted wave and evaluate the material samples scattering parameters. Particularly in free-space setups, care must be taken that the sample is large enough to prevent edge-effects such as diffraction around the specimen. Though it provides less accurate measurements on low-loss materials, a broadband characterization of the dielectric properties is possible. From the frequency dependent transmission/reflection behaviour, the relative permittivity and the loss tangent can be identified as functions of frequency, which is not directly possible in a resonant setup without changing the resonator itself. Furthermore, these methods allow the investigation of different layers in composite materials, if the layer thickness is within the bandwidth related resolution limit.

Additionally, open ended cable- or waveguide-structures that are contacted to the material under test as a probe can be used to identify the dielectric properties. The radiation characteristic into a homogenously filled half-space in front of the open-ended waveguide is subject to the presence of a material. By fitting a material-dependent model of the reflection behaviour to the measured characteristic, the material properties can be defined over the complete frequency range of the waveguide. However, care must be taken when choosing the sample size, since it should be thick enough to prevent multilayer behaviour.

## 5. Currently employed microwave dielectric materials

Polymeric and ceramic filled polymer composite materials have an important role to play in numerous microwave systems, both lying outside the scope of this review. Materials, including epoxy-derived products or polyimides, have long been used in many microwave technologies as substrates and filters. Their low, thermally stable permittivity renders them suitable also for various applications towards the higher frequency range of microwaves [63]. The low cost and facile formability of polymers mean that they will continue to play an important role in mmW applications [64]. However, there are notable drawbacks in the material performance of polymers that are becoming increasingly difficult to overcome in the development of emerging mmW devices. The low melting temperature of polymers prohibits their implementation in the processing of co-fired components as well as their use in adverse conditions or high temperature applications. Importantly, polymers are of limited use in mmW circuits due to their higher levels of dielectric losses and low $Q_f$ values, which are not comparable to those ultra-high values that have been realized in ceramics.

In the following, we focus on ceramics, addressing in more details two aspects – (i) low temperature co-firing as an example of currently employed fabrication technology and (ii) currently employed ceramic materials.

**Low temperature co-fired ceramics**

The diverse components used in high frequency applications are also commonly implemented in integrated circuits. They consist of an insulating substrate, on which dielectrics and conductive strips, often copper or gold, are deposited. For miniaturization and rapid synthesis, substrates and circuit components, including both conductive and dielectric materials, are often thermally processed together using multiple stacked layers of precursor materials connected by VIAs (vertical interconnection access). In order to simultaneously process ceramic and metallic materials together, low temperature co-fired ceramics (LTCC) are often employed. These LTCCs generally are required to sinter and densify at firing temperatures below 950°C, as this allows the use of conductive traces, e.g. Au, Ag or Cu. In addition to temperature stable dielectric properties, LTCC materials further need to be chemically compatible with the electrode material, have high thermal conductivity, low coefficient of thermal expansion (CTE), low toxicity, and of course, have a melting temperature above that of the electrode material [65]. The expansion of operating frequencies necessitates further decreases to dielectric losses in ceramics to maintain signal strength [66]. Although a great variety of low-loss ceramic materials have been reported in recent decades, most of them require rather high firing temperatures. The ongoing challenge in the field of LTCC materials is the design of materials requiring low sintering temperature without the addition of additives such as glasses that would indeed lower the sintering temperature but at the same time impair the dielectric performance.

An example of a component that is readily fabricated with LTCC technology are multilayer ceramic capacitors (MLCCs). Due to their fast charge-discharge behaciour, energy storage devices in the MLCCs are key constituents in high-power applications such as microwave and millimetre wave technologies, automobiles, especially hybrid electric vehicles, and telecommunication [67]. Their main advantages are small component sizes, wide capacitance and voltage ranges, and generally very stable performances over temperature. Typically, MLCCs are monolithic blocks of ceramic dielectric sheets interleaved with sheeted metal electrodes extending to opposite ends of the ceramic layers. To enable functionalities in very high frequency ranges such as impedance matching, decoupling and resonance circuits, MLCCs are being developed with larger numbers of layers and increased active area of overlapping electrodes. This increase, however, also requires more internal electrode material, typically Pt and Ag. In order to meet the ever increasing demand for MLCCs, cost reducing measures have to be taken by substituting precious metal electrodes with base metal electrodes (BMEs) such as Ni or Cu [68]. The use of BMEs in turn requires the ceramic sheets to have a lower sintering temperature than the melting point of the BME (1453°C for Ni), and the ceramic material must further be stable in a reducing atmosphere, which is necessary to avoid oxidation of the BME.

**Currently employed ceramic materials**

The most prominent, currently employed ceramic materials for commercial microwave dielectrics are complex perovskites of the general form $A(B1_{1/3}B2_{2/3})O_3$. Many studies on the dielectric performance in high frequencies for e.g. $Ba(Zn_{1/3}Ta_{2/3})O_3$ (BZT), $Ba(Zn_{1/3}Nb_{2/3})O_3$ (BZN) or $Ca_5TiNb_2O_{12}$ can be found in the literature [69–74]. Further perovskite materials, such as $BaTiO_3$ [75] or La-doped $Pb(Zr,Ti)O_3$ [76–78] have been subject of focused research towards ceramic layers in novel MLCCs, but due to their rather high permittivity values and necessary additives, they suffer from low quality factors in microwave frequencies which renders them unusable for high frequency and high precision BME-MLCCs [79]. Due to its flexible crystal structure, a multitude of metallic elements can be incorporated into the perovskite structure, allowing a wide range of tunability of $\varepsilon_r$ in the range between 30 and 90. The major drawback of this group of materials is, however, that very few compositions with ultra high $Q_f$ values, exceeding 200 000 GHz, have been reported so far. Another limitation is that because of perovskite's crystal chemistry, no $\varepsilon_r$ values lower than 20 are realisable. For the practical implementation as mmW materials, components with these permittivities have to be extraordinarily small (see Figure 5) which in turn requires more complicated production pathways via expensive micromachining [80]. Additionally, sintering temperatures of most ternary perovskites exceed those for LTCC technologies, rendering them unsuitable for prospective substrate applications. Lastly, because of the comparably high cost and scarcity of tantalum and niobium, there is a need to identify alternative materials based on abundant alkaline earth or base metals.

For millimetre wave applications, including waveguides and resonators, $Al_2O_3$ (alumina) ceramics have been the subject of extensive research. $\varepsilon_r$ values in alumina ceramics tend to be rather low, but increase with temperature (corresponding to a negative $\tau_f$), as do dielectric losses, making the use of this material problematic in mmW components. The best results in terms of dielectric properties have been achieved by adding small amounts of $TiO_2$, which has a strongly positive $\tau_f$ [81,82] and by additional doping with MnO to obtain exceptional loss characteristics [83]. One major drawback of $Al_2O_3$ - $TiO_2$ composites, however, is the formation of an intermediate $Al_2TiO_5$ phase, which has a strongly negative $\tau_f$ and deteriorates the ceramic's dielectric performance. To counteract the formation of $Al_2TiO_5$, sintering is carried out under GPa pressures [81], or an annealing step is applied following sintering to eliminate the secondary phase [82]. Pure alumina ceramics have further been evaluated as additively manufactured resonators and 2D photonic crystals showing great layout flexibility. However, high sintering temperatures (1600°C) and poorer dielectric performance compared to $Al_2O_3$ - $TiO_2$ composites limit the applications of such materials [84].

Towards millimetre wave and lower frequency microwave applications, further impactful research contributions exist for ilmenite and spinel structure type materials such as the binary titanates of $MgTiO_3$, $CaTiO_3$, $(Mg_{0.95}Co_{0.05})TiO_3$, $(Mg_{0.95}Zn_{0.05})TiO_3$, $Mg_2TiO_4$, and $(Mg_{1-x}Zn_x)_2TiO_4$ combining good loss characteristics with compositionally dependent medium to high levels of permittivity, which tend to increase with temperature due to negative $\tau_f$ values [85–89]. Although first commercially used DRs for the miniaturization of components in microwave circuits were based on high dielectric constant materials such as tungsten bronze-type $Ba_{6-3x}R_{8+2x}Ti_{18}O_{54}$ (R = rare earth metals) compounds [90], these compounds will most probably not be utilized anymore in technologies emerging in the high GHz range due to their low ceiling in terms of attainable loss characteristics.

Glass-ceramics find widespread use in industry as LTCC materials. Here, silicate-based glass-ceramics are already implemented in many commercially available products either as main material or as additives to certain LTCC compositions to lower their processing temperatures. State of the art was impressively summarized by Sebastian et al. [14]. Additionally, low permittivity silicate LTCCs have been the subject of several recent studies, which have identified that while dielectric properties may be attractive for high frequency applications, further work is required to reduce processing temperatures [91,92]. Of course, lowering processing temperatures is not only desirable in the context of co-firing, but universally appreciated for any component as it reduces cost and saves energy. Several studies with the intention of achieving ultra-low processing temperatures for silicates have been further conducted using cold sintering methods [93–96], with promising results. Also, a study summarizing the issues and potentialities for extremely low temperature methods for ceramics has been recently published [97]. Utilising silicates as high Temperature Co-fired Ceramics (HTCCs) or in the context of cold sintering techniques is expected to present lower barriers to densification and has drawn increasing interest in recent years [94,98]. The ongoing developments of new techniques in the fields of HTCCs and cold sintering are expected to be instrumental in enabling the utilisation of silicate dielectric ceramics in future mm wave applications.

As yet, silicate ceramics find only limited commercial use in high frequency applications, primarily as integration and substrate solutions. At currently harnessed frequencies, classical oxide ceramic and polymeric materials just barely meet the requirements in terms of quality factors and due to their completed and optimised development in current technologies they remain the dominant dielectric materials. However, the introduction of new technologies using ever high frequencies will create a demand for lower loss materials, which silicate ceramics could fulfill. Due to the attractiveness of their properties towards high frequency dielectric performance, silicates from Neso- and Sorosilicate families as DRs have been proposed by several studies in recent years [18,99]. Further, the use of silicates in the context of high frequency MLCCs presents an attractive pathway towards low-cost dielectric components with reduced environmental impact and motivates the further development of new materials and methods.

**Material limitations and paradigm shift**

Although the development of novel mmW dielectrics has been the subject of much research in recent years, a few key material limitations endure:

- although great advances in terms of materials with either very low $\varepsilon_r$, ultra high $Q_f$ or near zero $\tau_f$ have been made, there is still a lack of materials that combine all three of those requirements

- there is a huge demand for materials with low sintering temperatures that are co-firable with low melting temperature metal electrodes for implementation in LTCCs and MLCCs. It is of particular importance to develop materials with low sintering temperatures without the addition of flux additives such as glasses, as these will deteriorate the desired dielectric properties described above
- New materials should be derivable from abundant resources. This drive is naturally generated by the desire for low cost production and environmental responsibility, as well as potential supply bottlenecks concerning scarce elements such as Hafnium, Tantalum, Niobium or rare earth elements

## 6. Silicate materials

Silicates are often described as crystalline materials in which silicon is tetrahedrally coordinated by oxygen. It is not surprising that silicates are the major rock forming minerals on all terrestrial planets in our solar system, considering that silicon is the second most terrestrially abundant element and the chemical bond between Si and O is stronger than with any other element [100]. In fact, more than 90% of all minerals in Earth's crust are either feldspars, pyroxenes, amphiboles, micas or quartz, all of which are silicates. Most of these minerals don't have a fixed composition but are highly diverse in their chemical make-up, allowing for a great range of substitutions and structural modifications.

Based on a purely electrostatic approach of describing silicates, oxygen ions would form a close-packed sublattice structure with silicon and other cations occupying interstitials. However, a more appropriate model to describe the Si-O bond is that of covalent bonding according to the octet rule with overlapping orbitals. This description is in good agreement with the tetrahedral coordination of Si by oxygen in silicate materials. In reality, a combination of both models is accepted with the total Si-O bonding character of 55% covalent and 45 ionic% [101]. The covalent nature of bonding in silicates is one of the distinguishing features of this class of oxides.

The ubiquitous $SiO_4$ unit in silicates is the foundation for classifying silicate minerals. Generally, $SiO_4$ tetrahedra are connected with each other via their apices, forming the mineral's framework with interstices occupied by other cations. Arranging the $SiO_4$ units in different ways results in different frameworks with varying tetrahedral connectivity. The degree of polymerization extends from isolated tetrahedra on the one extreme, to a 3D network in which all tetrahedra share all four oxygens with neighbouring tetrahedra, on the other end. A complete classification is given in Table 2.

To explain its compositional diversity, it is important to understand the geometric flexibility of silicate structures. The strength of the Si-O bond limits its length to 1.60 – 1.64 Å, depending on the coordination number of oxygen. The real rigidity of silicate frameworks arises from the pliability of the Si-O-Si bond angle in corner-shared tetrahedra ranging from 120° and 180° [102]. The flexibility of this angle allows the silicate structure to react to distortions, e.g. by the introduction of different cations of varying sizes or temperature and pressure changes, by expansion or contraction with relative ease, thus maintaining its rigid skeleton.

**Table 2: Different silicate structures based on their simplest building unit.**

| | 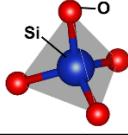 | | | |
|---|---|---|---|---|
| **Type** | **Structure** | **SiO$_4$-linkage** | **Main building unit** | **Silicate dielectric material example** |
| Neso- | 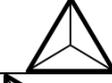 | isolated | $[SiO_4]^{4-}$ | Forsterite Mg$_2$SiO$_4$ |
| Soro | 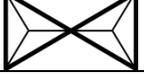 | Double tetrahedra | $[Si_2O_7]^{6-}$ | Åkermanite Ca$_2$MgSi$_2$O$_7$ |
| Cyclo | 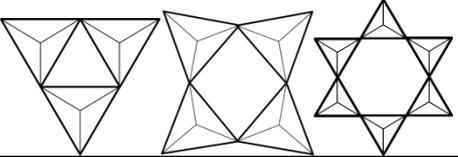 | 3-,4- or 6-rings | $[Si_3O_9]_n^{6n-}$ | Cordierite/Indialite Mg$_2$Al$_4$Si$_5$O$_{18}$ |
| Ino | 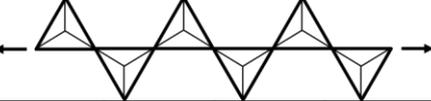 | Single chains | $[SiO_3]_n^{4n-}$ | Enstatite Mg$_2$Si$_2$O$_6$ |
| Ino | 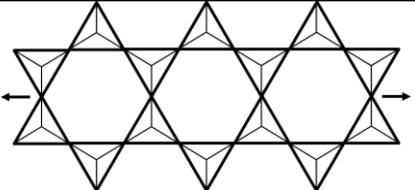 | Double chains | $[Si_4O_{11}]_n^{6n-}$ | none known |
| Phyllo | 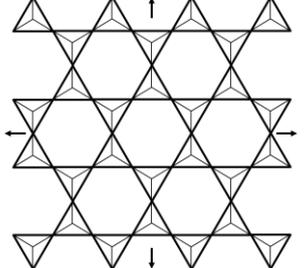 | Sheets | $[Si_2O_5]_n^{2n-}$ | Gillespite SrCuSi$_4$O$_{10}$ |
| Tecto | 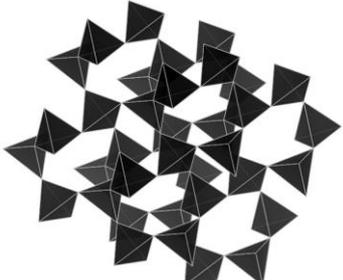 | 3D network | $[SiO_2]_n$ | Celsian BaAl$_2$Si$_2$O$_8$ |

**Suitability of silicates as mmW dielectric materials**

Here, we briefly discuss why silicate ceramics fulfill the requirements for millimetre wave materials described in section 3. Silicate ceramics are synthetic materials that tend to exhibit crystalline structures analogous to those found in naturally occurring minerals. By controlling synthesis conditions and compositions, a broad range of silicate ceramic materials can be fabricated with diverse physical properties.

To address the increasingly demanding material requirements of emerging millimetre wave applications, such silicate ceramics present attractive pathways to the design of high performance mmW dielectrics. In the following, we will highlight their attractive features.

**Low permittivity**

As discussed earlier, the extension of operating frequencies of e.g. 5G and Radar applications requires dielectric materials with $\varepsilon_r$ values <10 according to Equation 3. Silicates are promising candidates in the field because of the crystal chemical peculiarity of the $SiO_4$ tetrahedron. The Si-O bonds in $SiO_4$ tetrahedron have a bonding character of only 45% ionic and 55% covalent [101], leading to very low dielectric polarizability values and thus a decreased rattling effect and very low relative permittivity values [10]. This is in contrast to many other oxides, which tend to be predominantly ionic in their bonding character. For example, in perovskites of the titanate family, the Ti-O bond in the structure's main building block, the $TiO_6$-octahedron, is of almost pure ionic character, which in turn is conducive to ionic displacement in response to an electric field, resulting in higher dielectric polarizabilities and permittivities.

**Low temperature coefficients**

To interpret the relatively low temperature drift of resonating frequencies for silicate materials, it important to understand the origin of $\tau_f$. It is clear from Equation 11, that, because thermal expansion values in electronic ceramics are usually very low, the value of $\tau_f$ is governed predominantly by $\tau_\varepsilon$. The following equation is derived from the CME and describes the origin of $\tau_\varepsilon$:

$$\tau_\varepsilon = \frac{(\varepsilon_r - 1)(\varepsilon_r + 1)}{\varepsilon_r}\left[\frac{V_m}{\alpha_D}\left(\frac{\delta\alpha_D}{\delta V_m}\right)_T \alpha_l + \frac{1}{3\alpha}\left(\frac{\delta\alpha_D}{\delta T}\right)_{V_m} - \alpha_l\right] \quad (12)$$

The comparably low bond ionicity of silicates described above leads to a decrease in the right term of this equation and to overall lower $\tau_\varepsilon$ values for silicates. Wersing et al. [103] showed that it is indeed only covalent inorganics such as $SiO_2$ that possess the combination of $\varepsilon_r$, $\tau_\varepsilon$ and $\alpha_l$ close to 0 out of all discussed high frequency dielectric materials.

**Dielectric loss mechanisms and quality factors in silicates**

High quality factors are a result of low dielectric losses in dielectric materials. Although it has been shown that material's atomic arrangement and its manipulation has effect on its dielectric loss, it is usually the macroscopic defects of the bulk material arising from imperfect processing that are the dominating factors in the loss of a dielectric component. In the following, we will discuss both of these factors and their effect on the quality factor in detail.

Dielectric losses are an unavoidable consequence of the interaction of electromagnetic waves with materials at high frequencies. Lost energy dissipates into heat, increasing the temperature of the components. Losses arise from mechanisms that can be split into two categories: (i) intrinsic factors that are characteristic of the materials crystal lattice and are therefore a material property and (ii) extrinsic factors that arise from material features at the microstructural level, such as grain boundaries, pores, strain fields or impurities. Such features or imperfections are chiefly related to a given material's processing rather than composition.

Intrinsic loss can be considered as the remaining loss in energy in an ideal dielectric material as a result of the interaction of the phonon system with the electric field and therefore allows an estimation of the lower limit of loss in a perfectly crystalline material. It is therefore not surprising that the lowest known levels of microwave losses are found in single crystalline materials [18,104,105] that can approach to this lower limit of loss. Fundamentally, the intrinsic loss can be described as a collision process between the electromagnetic field and thermal phonons in which energy quanta are absorbed [106]. Intrinsic loss is symmetry sensitive, and the differences in the intrinsic loss in centro- and non-centrosymmetric materials are governed by different quantum and quasi-Debye contributions [107]. This was demonstrated by several authors [108–111] for substituted $Mg_2Al_4Si_5O_{18}$ ceramics, in which the transformation from cordierite to indialite polymorphs was accompanied by an increase in symmetry and significantly higher Q-values. Additionally, Ferreira et al. [112] showed that the introduction of foreign atoms into the crystal lattice by doping could lead to anharmonic interactions and increased intrinsic loss. In addition to increasing the value of $\varepsilon_r$, high polarizability is usually also disadvantageous in terms of Q, since this leads to easy displacement and movement of ions inside the lattice. Ohsato et al. formulated synoptically that a fundamentally ideal microwave dielectric is of high symmetry, with an inversion symmetry element and no defects in the lattice [113].

Extrinsic losses originate from material features stemming from synthesis or processing related factors. In a perfect single crystal, dielectric losses would be limited to those intrinsic factors described above. However, when preparing ceramic samples, a multitude of parameters influencing the obtained microstructure have to be considered. Less than perfect densification will always lead to some degree of porosity in the final material while doping or substitution can lead to secondary phase formation and impurities. In any case, for polycrystalline ceramic materials, grain boundaries are inevitably present in the product material. Structural disruption at such defect sites creates local dipoles, that increase total overall polarizability of the material which in turn decreases Q [9,114].

Although it is impossible to eliminate extrinsic losses completely, especially in ceramics, one can take measures to limit their extent. Optimizing the sintering behaviour to obtain the highest degree of densification of the ceramic can reduce porosity to a minimum and reduce extrinsic loss [15][115]. This can be achieved by, e.g., cyclic pre-pressing, calcination and regrinding of the powder starting material, and optimal sintering temperatures. Further, the presence of already very small amounts of secondary phases (<1%) was shown to have a large negative effect on Q in barium zinc tantalate ceramics [70,71]. In a similar manner, to obtain low-loss product ceramics, liquid phase formation during sintering should be avoided since it deteriorates densification and increases dielectric loss [116–118]. In terms of sintering duration, the correct procedure for a material is not straightforward. On the one hand, it was demonstrated that in some cases, fast heating with short sintering at a temperature higher than an order-disorder transition for a material could significantly increase the Q value of a ceramic [110,119]. Here the authors argue that the reason for that is increased symmetry in disordered structures. Opposing these findings, Surendran et al. reported that a slight non-stoichiometry-driven ordering of cation sites increases Q-values [116]. It is noteworthy that both of these studies were performed on perovskite materials. The latter findings are supported by Schlömann [120], who concluded that ion disordering increases loss when it violates periodicity. On the other hand, if there is no order-disorder transition, longer sintering time has also been shown to increase Q, due to the reduction of grain boundaries with increasing grain growth [121,122]. However, sintering for an excessively long duration may lead to abnormal grain growth, which will in turn, decrease Q again [123]. A slow cooling rate is recommended when Si is substituted with Al in a 1:1 ratio since a long-range equilibrium order of with alternating Al and Si inside the tetrahedral can be introduced, increasing the Q values in feldspars. [124,125]

Grain boundaries are generally seen as one of the main contributors to high frequency loss in ceramics. In their study from 2009, Breeze et al. consider the effect of grain boundaries to be mainly as a sink for impurities and write them off as the cause for any influence on microwave dielectric loss [126]. Lastly, the choice of starting materials has been proven to influence the quality of the products immensely. Using nanosized ceramic precursors can lower the sintering temperature and have a positive effect on Q [127–129]. This approach was taken further by Ando et al. by using amorphous starting material and improving the loss characteristics of forsterite ceramics [130]. Choi et al. concluded recently [131], that by adding only 0.5 wt% of $Cr_2O_3$ to a glassy $CaMgSi_2O_6$ reduces its nucleation temperature significantly and yields ceramics with improved crystallinity and dielectric performance. Another study came to the conclusion that it is actually the dispersion characteristics that have a larger effect on the product ceramics than the particle size of the starting powders [132]. In a recent study Du et al. [133] concluded that besides greater dielectric polarizability, a higher packing fraction, given as

$$packing\ fraction = \frac{volume\ of\ the\ atom\ in\ the\ cell}{volume\ of\ unit\ cell} * Z \quad (13)$$

also corresponds to higher quality factor values in apatites of the form $CaRE_4Si_3O_{13}$. In the same work, the authors also demonstrate a direct proportionality between polyhedral distortion and nearer to zero values of $\tau_f$ in the same compounds.

A survey of the literature in terms of processing parameters reveals inconsistencies in many cases. Although it is clear that pores and liquid phase formation are generally undesirable for low loss materials, many contradicting observations have been made regarding the role of grain size, starting materials, heating and cooling regimes or crystal chemical factors on the extrinsic dielectric loss of ceramics.

**Cation substitution and tunability of dielectric properties**

As described earlier, frameworks of all silicate families can accommodate a broad range of cation substitutions. As a result, an exceptionally high number of complete solid solutions between silicate end member compositions can be observed in both natural and synthetic varieties. An example is the olivine group in which the endmembers Forsterite ($Mg_2SiO_4$), Fayalite ($Fe_2SiO_4$), and Larnite ($Ca_2SiO_4$) form a triangle of complete solubility. Similar to the response of the lattice parameters to the substitution according to Vegard's law [134], cation substitution is directly coupled to a change in dielectric polarizability and susceptibility according to Equations 5 and 8 of the respective compound. This dependence gives rise to a predictive tunability of $\varepsilon_r$ and $\tau_f$ values which is of great importance for components with restricted dimension in a defined operating frequency. Although cation substitution in silicates is most easily realised for metal cations in the usually octahedral environment, also coupled substitutions involving the $SiO_4$ tetrahedra are possible. A prominent example are the feldspar solid solutions between Albite - Anorthite ($NaAlSi_3O_8$ – $CaAl_2Si_2O_8$) in which the $Na^+$ - $Ca^{2+}$ exchange is balanced by an $Si^{4+}$ – $Al^{3+}$ exchange on the tetrahedral site. This type of substitution involving two different cation sites is often termed Tschermak-exchange [135] and very common in silicates.

In summary, silicates represent potent material candidates for implementation in high frequency technologies because of their unique combination of low $\varepsilon_r$ and near zero $\tau_f$ values and extensive possibilities to tune the relevant properties by substitution. The lower limit of dielectric loss for a dielectric material can be estimated by calculating the remaining loss in a perfectly crystalline material due to intrinsic phenomena. However, it is usually the extrinsic loss mechanism that determines the scale of the $Q_f$ values of the bulk material. It is therefore hard to pinpoint the net loss for any given ceramic dielectric material without any experimental work, and to achieve an optimum it should always be the focus to improve synthesis, densification and thermal treatment. The attainability of ultra high $Q_f$ values in silicate ceramics of all families has been repeatedly demonstrated. Quality factors in silicate ceramics are mostly well above the values of known perovskite and polymeric materials. A comprehensive survey of literature relating to the dielectric performance of silicate ceramics was conducted here. In Table 3 we present a summary of dielectric performance parameters and sintering conditions reported for silicates from a range of studies. When we combine the reported data into a single figure to assess trends and identify future research directions, we can clearly observe how silicate ceramics fill the void for materials with low permittivity and ultra high Q factors necessary for millimetre wave applications (Figure 6).

Table 3: A summary of reported dielectric properties and sintering conditions for silicate ceramics.

| | Material | Sintering T (°C) | $\varepsilon_r$ | Qxf (GHz) | $f_0$ (GHz) | $\tau_f$ (ppm/K) | Ref. |
|---|---|---|---|---|---|---|---|
| | $Mg_2SiO_4$ from amorphous $SiO_2$ (0.25 µm spherical) | 1450, 2h | 6.9 | 219 200 | 15 | -68 | [130,136] |
| | $Mg_2SiO_4$ from amorphous $SiO_2$ (0.82 µm irregular) | 1400, 2h | 6.6 | 150 000 | 15 | -66 | [130,136] |
| | $Mg_2SiO_4$ | 1360, 2h | 6.9 | 240 000 | 23 | -67 | [117] |
| | $Mg_2SiO_4$ | 1400, 10h | | 250 000 | 15 | | [132] |
| | $Mg_2SiO_4$ | 1550, 3h | 7.5 | 114 730 | 10 | -59 | [137] |
| | $Mg_2SiO_4$ | 1500 | 5.1 | 2 333 | 7 | 114 | [138] |
| | $Mg_2SiO_4$ + 1 wt% $TiO_2$ | 1200, 2h | 7.0 | 230 000 | 17 | -65 | [139] |
| | $Mg_2SiO_4$ + 24 wt% $TiO_2$ | 1200, 2h | 11.0 | 85 000 | | 0 | [140] |
| | $Mg_2SiO_4$ + 6.5 wt% $TiO_2$ (infiltrated via LPD) | 1400, 2h | 5.5 | 10 500 | | -46 | [141] |
| | $Mg_2SiO_4$ + 15 wt% Li-Borosilicate glass | 950 | 5.0 | 1 000 | 7 | 277 | [138] |
| | $0.975(Mg_2SiO_4) – 0.025(Mn_2SiO_4)$ | 1400, 2h | 6.71 | 180 000 | | -75 | [142] |
| | $0.93(Mg_2SiO_4) – 0.07(Ca_2SiO_4)$ | 1400, 2h | 6.87 | 105 000 | | -73 | [142] |
| | $Mg_{1.5}Co_{0.5}Si_2O_6$ | 1350, 9h | 8.11 | 145 846 | | -12 | [143] |
| | $Mg_{1.92}Cu_{0.08}SiO_4$ | 1250, 4h | 6.35 | 188 500 | 10-16 | -2 | [144] |
| | $Mg_{1.9}Ni_{0.1}SiO_4$ | 1500, 4h | 6.8 | 152 300 | | | [145] |
| | $Mg_{1.9}Ni_{0.1}SiO_4$ + 12 wt% $Li_2CO_3-V_2O_5$ | 1150, 4h | 6.9 | 99 800 | | -50 | [145] |
| | $Mg_2(Si_{0.9}Ti_{0.1})O_4$ | 1350, 3h | 7.4 | 73 760 | 15 | -60 | [146] |
| | $Zn_2SiO_4$ | 1340 | 6.6 | 219 000 | | -61 | [147] |
| | $Zn_2SiO_4$ | 1325, 4h | 6.6 | 198 400 | 14-16 | -41.6 | [148] |
| | $Zn_2SiO_4$ | 1250, 3h | 6.7 | 27 917 | | -17 | [121] |
| | $Zn_2SiO_4$ + 20 wt% Zn-Borosilicate glass | 900 | 6.85 | 31 690 | | -28.2 | [149] |
| | $Zn_2SiO_4$ + 12 mol% $V_2O_5$ | 875, 2h | 7.3 | 17 500 | 20 | -28 | [150] |
| | $Zn_2SiO_4$ + 11 wt% $TiO_2$ | 1200, 4h | 9.1 | 150 800 | 14-16 | -1 | [148] |
| | $Zn_2SiO_4$ + 11 wt% $TiO_2$ | 1250 | 9.3 | 113 000 | | 1 | [147] |
| | $ZnSiO4$ + 10 wt% nano-sized $TiO_2$ | 1275, 5h | 8.9 | 37 902 | 13.8 | -9.6 | [151] |
| N e s o | $Zn_2SiO_4$ + 5 wt% $Li_2CO_3$ + 4 wt% $Bi_2O_3$ | 910, 2h | 6.65 | 33 000 | 11 | -70 | [152] |
| | $Zn_{1.6}Mg_{0.4}SiO_4$ | 1170, 2h | 6.3 | 189 800 | | -63 | [153] |
| | $Zn_{1.2}Mg_{0.8}SiO_4$ | 1250, 3h | 6.6 | 95 650 | 10 | -60 | [154] |
| | $Zn_{1.95}Co_{0.05}SiO_4$ | 1400, 6h | 6.35 | 225 198 | | -23 | [155] |
| | $Zn_{1.8}SiO_{3.8}$ | 1300, 3h | 6.6 | 147 000 | | -22 | [121] |
| | $Zn_{1.7}SiO_{3.7}$ + 25 mol% $B_2O_3$ | 900, 2h | 6.1 | 70 000 | | -21.9 | [156] |
| | $Zn_{1.9}Co_{0.1}SiO_4$ + 2 wt% $Li_2O-B_2O_3-SiO_2-CaO-Al_2O_3$ glass | 900, 3h | 6.5 | 57 000 | 20 | -55 | [157] |
| | $LiAlSiO_4$ | 1350, 10h | 4.8 | 36 000 | | 8 | [158] |
| | $LiAlSiO_4$ + 12 mol% $B_2O_3$ | 950, 10h | 5.3 | 212 000 | | -7.7 | [158] |
| | $Li_2MgSiO_4$ | 1100, 2h | 5.73 | 13 559 | 8 | | [159] |
| | $Li_2MgSiO_4$ | 1250, 2h | 5.1 | 15 285 | 8 | | [160] |
| | $Li_2MgSiO_4$ + 0.1 wt% $B_2O_3$ | 1100, 4h | 5.93 | 9 523 | 8 | | [159] |
| | $Li_2MgSiO_4$ + 0.5 wt% $MgF_2$ | 1100, 4h | 5.75 | 10 959 | 8 | | [159] |
| | $Li_2MgSiO_4$ + 0.5 wt% $WO_3$ | 1100, 4h | 6.03 | 12 121 | 8 | | [159] |
| | $Li_2MgSiO_4$ + 1 wt% Li-Borosilicate glass | 925, 2h | 5.5 | 114 286 | 8 | | [160] |
| | $Li_2MgSiO_4$ + 2 wt% Li-Mg-Zn-Borosilcate glass | 875, 2h | 5.9 | 119 403 | 8 | | [160] |
| | $Li_2Mg_{0.6}Zn_{0.4}SiO_4$ + 3 wt% Li-Mg-Zn Borosilicate glass | 900, 3h | 5.89 | 44 787 | | -71.65 | [161] |
| | $Li_2Mg_{0.95}Co_{0.05}SiO_4$ | 1050, 4h | 5.75 | 26 500 | 16 | -9.4 | [162] |
| | $Li_2Ca_{0.96}Mg_{0.04}SiO_4$ | 925, 4h | 5.63 | 4 210 | 8 | | [163] |
| | $Li_2TiSiO_5$ | 1180, 6h | 9.89 | 38 100 | 14.2 | 50.1 | [164] |
| | $ZrSiO_4$ | 1550, 4h | 7.4 | 85 000 | 5.2 | | [165] |
| | $HfSiO_4$ | 1600, 2h | 7.0 | 25 000 | 10 | -44 | [166] |
| | $Al_2SiO_5$ | 1525 | 4.43 | 41 800 | | -17 | [94] |
| | $Al_2SiO_5$ | 1500, 4h | 4.5 | 6 250 | 5 | | [98] |
| | $Al_2SiO_5$ + 50 wt% NaCl | 120 | 4.52 | 22 350 | | -24 | [94] |

| | Material | | | | | | |
|---|---|---|---|---|---|---|---|
| | $Sm_2SiO_5$ | 1500, 4h | 8.44 | 64 000 | | -37 | [167] |
| | $Nd_2SiO_5$ | 1500, 4h | 7.94 | 38 800 | 18.4 | -53 | [168] |
| | $CaSnSiO_5$ | 1525, 4h | 9.08 | 61 000 | | 35 | [169] |
| | $CaSnSiO_5$ | 1450, 5h | 11.4 | 45 000 | 9-12 | 62.5 | [170] |
| | $Ca_{1.8}SnSi_{1.4}O_{6.4}$ | 1450, 5h | 10.2 | 81 000 | 9-12 | -4.8 | [170] |
| | $Ca_3MgSi_2O_8$ | 1375, 3h | 13.8 | 27 000 | 10.8 | -62 | [171] |
| | $CaNd_4Si_3O_{13}$ | 1350, 5h | 14.8 | 16 300 | 11-13 | -18.3 | [133] |
| | $CaEr_4Si_3O_{13}$ | 1400, 5h | 14.1 | 18 600 | 11-13 | -17.8 | [133] |
| | $SrY_2Si_3O_{10}$ | 1425, 4h | 9.3 | 64 100 | | -31 | [172] |
| | $BaY_2Si_3O_{10}$ | 1400, 4h | 9.5 | 65 600 | | -28 | [172] |
| | $Sr_2ZnSi_2O_7$ | 1475, 2h | 8.4 | 105 000 | 12.63 | -51.5 | [173] |
| | $Sr_2ZnSi_2O_7 + 2\ wt\%\ SrTiO_3$ | 1450, 2h | 8.8 | 60 000 | 12.5 | -13 | [173] |
| | $Sr_2ZnSi_2O_7 + 15\ wt\%\ Li\text{-}Mg\text{-}Zn\text{-}Borosilicate\ glass$ | 875, 2h | 7.9 | 39 000 | 12.6 | -53.5 | [122] |
| | $Sr_2CoSi_2O_7$ | 1375, 2h | 8.9 | 34 000 | 10-18 | -56.7 | [173] |
| | $Sr_2MgSi_2O_7$ | 1375, 2h | 8.3 | 55 000 | 10-18 | -47.5 | [173] |
| | $Sr_2MnSi_2O_7$ | 1375, 2h | 8.8 | 32 000 | 10-18 | -58.8 | [173] |
| | $Sr_2Al_2SiO_7$ | 1525, 4h | 7.0 | 25 000 | 12-16 | -35 | [174] |
| | $SrCaAlSiO_7$ | 1475, 4h | 8.4 | 25 000 | 12-16 | -42 | [174] |
| | $Sm_2SiO_2O_7$ | 1375, 2h | 10.0 | 1 587 | 10 | 63 | [175] |
| | $Sm_2SiO_2O_7 + 15\ wt\%\ Li\text{-}Borosilicate\ glass$ | 975, 2h | 9.89 | 4 000 | 10 | | [175] |
| S o r o | $Sm_2SiO_2O_7 + 15\ wt\%\ Li\text{-}Mg\text{-}Zn\text{-}Borosilicate\ glass$ | 975, 2h | 9.09 | 3 846 | 10 | | [175] |
| | $Ba_2MgSi_2O_7$ | 1350, 10h | 7.0 | 30 000 | | -62 | [176] |
| | $BaCo_2Si_2O_7$ | 1060, 3h | 9.26 | 31 135 | 12-14 | -92.05 | [177] |
| | $BaCu_{1.85}Co_{0.15}Si_2O_7$ | 1000, 3h | 8.45 | 58 958 | 12-14 | -34.4 | [178] |
| | $Ba_2ZnSi_2O_7$ | 1300, 10h | 7.0 | 48 000 | | -74 | [176] |
| | $Ba_2ZnSi_2O_7$ | 1200, 3h | 8.09 | 26 600 | 12.5 | -51.4 | [60] |
| | $Ba_2ZnSi_2O_7 + BaSiO_3$ | 1125, 3h | 8.3 | 27 200 | 12.5 | -41.5 | [60] |
| | $Ca_2MgSi_2O_7$ | 1300, 4h | 9.86 | 8 016 | -42 | 7.9 | [179] |
| | $Ca_2ZnSi_2O_7$ | 1300, 2h | 11.0 | 13 500 | 10-18 | -64.3 | [173] |
| | $Ca_2Al_2SiO_7$ | 1500, 4h | 9.3 | 31 000 | 12-16 | -34 | [174] |
| | $Ca_2Al_2SiO_7$ | 1440, 4h | 8.86 | 22 457 | | -51.06 | [180] |
| | $Ca_3SnSi_2O_9$ | 1400, 10h | 8.4 | 93 296 | 12-15 | -64.7 | [181] |
| | $Ca_3SnSi_2O_9$ | 1500, 4h | 8.44 | 92 000 | 15.1 | -60 | [181] |
| | $Ca_3ZrSi_2O_9$ | 1400, 10h | 10.6 | 54 791 | 12-15 | -76.8 | [181] |
| | $Mg_2Si_2O_6$ | 1380, 13h | 6.7 | 121 200 | | -17 | [182] |
| | $Mg_{1.7}Zn_{0.3}Si_2O_6$ | 1360, 9h | 8.05 | 138 481 | | -14 | [123] |
| | $Mg_{1.8}Ni_{0.2}Si_2O_6$ | 1425, 9h | 6.1 | 118 702 | | -10 | [183] |
| | $Mg_{1.8}Ni_{0.2}Si_2O_6 + 3\ wt\%\ TiO_2$ | 1300, 9h | 8.29 | 101 307 | | -2.98 | [184] |
| | $Mg_{1.8}Ca_{0.2}Si_2O_6$ | 1330, 2h | 7.45 | 47 500 | | -47 | [185] |
| | $CaMgSi_2O_6$ | 1300, 2h | 7.6 | 121 381 | | -66 | [186] |
| | $CaMgSi_2O_6$ | 1290, 2h | 7.46 | 59 638 | | -46 | [185] |
| | $CaMgSi_2O_6 + 12\ wt\%\ Al_2O_3$ | 1250, 2h | 7.99 | 60 132 | | -48.21 | [58] |
| | $CaMgSi_2O_6 + 0.5\ wt\%\ Cr_2O_3$ | 900, 1h | 6.98 | 50 460 | 12 | -23 | [131] |
| | $CaMgSi_2O_6 + 15\ wt\%\ Li\text{-}Borosilicate\ glass$ | 925, 2h | 8.0 | 15 000 | 10.17 | -49 | [187] |
| | $CaMgSi_2O_6 + 15\ wt\%\ Li\text{-}Mg\text{-}Zn\text{-}Borosilicate\ glass$ | 900, 2h | 8.2 | 32 000 | 10.15 | -48 | [187] |
| | $CaMgSi_2O_6 + 12\ wt\%\ CaTiO_3 + 1\ wt\%\ Li_2CO_3+V_2O_5$ | 880, 2h | 9.23 | 46 200 | | 1.3 | [188] |
| | $CaMg_{0.96}Cu_{0.04}Si_2O_6$ | 1250, 4h | 7.4 | 160 100 | | -42 | [189] |
| | $CaMg_{0.98}Mn_{0.02}SiO_6$ | 1300, 3h | 8.01 | 83 469 | 6-9 | -45.72 | [190] |
| | $CaMg_{0.9}Zn_{0.1}Si_2O_6$ | 1200, 4h | 7.88 | 76 100 | | -22.5 | [191] |
| | $CaMg_{0.9}Zn_{0.1}Si_2O_6 + 0.6\ wt\%\ LiF$ | 900, 4h | 7.69 | 69 900 | | -24.9 | [191] |
| I n o | $(CaMg_{0.92}Al_{0.08})Si_{1.92}O_6$ | 1275, 2h | 7.89 | 59 772 | | -42.12 | [192] |
| | $(CaMg_{0.92}Al_{0.08})Si_{1.92}O_6 + 22\ wt\%\ TiO_2$ | 1225, 2h | 12.06 | 5 940 | | 2.24 | [57] |
| | $(CaMg_{0.92}Al_{0.08})Si_{1.92}O_6 + 22\ wt\%\ TiO_2$ | 1250, 2h | 9.6 | 32 590 | | -1.53 | [57] |
| | $(CaCo_{0.92}Al_{0.08})Si_{1.92}Al_{0.08}O_6$ | 1150, 2h | 8.46 | 60 125 | 12.65 | -49.26 | [193] |
| | $(CaMg_{0.82}Co_{0.1}Al_{0.08})Si_{1.92}Al_{0.08}O_6$ | 1250, 2h | 7.59 | 60 535 | 12.72 | -42.17 | [193] |
| | $CaCoSi_2O_6$ | 1175, 2h | 6.04 | 12 457 | 5-15 | -18.91 | [194] |
| | $BaCuSi_2O_6$ | 1050, 3h | 8.45 | 31 424 | 12-14 | 0.36 | [195] |
| | $BaCuSi_2O_6 + 2\ wt\%\ LiF\text{-}MgF_2\text{-}SrF_2\ composite$ | 875, 3h | 8.16 | 24 351 | 12-14 | -9.74 | [196] |
| | $Ba_{0.8}Sr_{0.2}CuSi_2O_6$ | 1000, 3h | 8.25 | 47 616 | 12-14 | -9.61 | [195] |
| | $CaSiO_3$ | 1300 | 8.4 | 16 000 | 10 | | [91] |
| | $CaSiO_3$ | 1320, 2h | 6.69 | 25 298 | | | [127] |
| | $CaSiO_3 + 8.6\ wt\%\ B_2O_3$ | 800 | 7.9 | 2 222 | 10 | | [197] |
| | $\alpha\text{-}CaSiO_3 + 2\ wt\%\ TiO_2$ | 1300, 2h | 7.9 | 16 491 | | 0.76 | [56] |
| | $CaSn_{0.1}SiO_{3.2}$ | 1350, 5h | 7.6 | 32 000 | 9-12 | -0.2 | [170] |
| | $Ca_{0.9}Mg_{0.1}SiO_3$ | 1290, 2h | 6.49 | 62 420 | | -43.3 | [185] |
| | $Mg_2Al_4Si_5O_{18}$ | 1430, 2h | 6.2 | 39 900 | | -24 | [111] |
| C y c l | $Mg_2Al_4Si_5O_{18} + 7\ wt\%\ Yb_2O_3$ | 1375 | 4.9 | 112 500 | 18 | | [198] |
| | $Mg_2Al_4Si_5O_{18} + 5\ \%\ CaTiO_3$ | 1400, 3h | 7.2 | 55 490 | 15.3 | -28.3 | [199] |
| | $Mg_2Al_4Si_5O_{18} + 5\ wt\%\ P_2O_5 + 5\ wt\%\ B_2O_3$ | 950 | 5.0 | 6 600 | 16 | -15 | [200] |
| | $0.9Mg_2Al_4Si_5O_{18} - 0.1TiO_2$ | 1400, 4h | 6.3 | 55 400 | 18 | -21 | [201] |
| | $Mg_{1.8}Ni_{0.2}Al_4Si_5O_{18}$ | 1400, 2h | 6.15 | 99 110 | | -24 | [108,111] |
| | $Mg_{1.95}La_{0.05}Al_4Si_5O_{18}$ | 1425, 3h | 6.7 | 78 500 | 14.3 | -22 | [109] |
| | $Mg_{1.8}Sm_{0.2}Al_4Si_5O_{18}$ | 1275, 3h | 6.8 | 24 800 | 13.2 | -17 | [109] |
| P | $SrCuSi_4O_{10}$ | 1100, 6h | 4.0 | 4 545 | 5 | | [202] |
| | $SrCuSi_4O_{10}$ + 5 wt% Li-Mg-Zn Borosilicate glass | 900, 6h | 5.0 | 2 632 | 5 | | [202] |
| Tecto | $SiO_2$ (low Cristobalite) | 1650, 3h | 3.81 | 80 400 | 22 | -16.1 | [203] |
| | $SiO_2$ (low Cristobalite) | 1500, 3h | 3.52 | 92 400 | | -14.5 | [128] |
| | $SiO_2$ amorphous | 1100, 5h | 3.38 | 44 300 | 21 | -14.4 | [204] |
| | $SiO_2$ amorphous, SPS | 1000, 3 min | 3.9 | 63 500 | 21 | -5.7 | [204] |
| | $SiO_2$ amorphous, melting | melting | 3.83 | 122 100 | 21 | -8.2 | [204] |
| | $0.945\ SiO_2$ (tridymite) - $0.055\ Li_2TiO_3$ | 1050, 3h | 3.2 | 10 180 | | 0.17 | [205] |
| | $0.8\ SiO_2 – 0.2\ B_2O_3$ | 1100, 3h | 3.56 | 70 600 | | -11.4 | [206] |
| | $CaAlSi_2O_8$ | 1400, 15h | 7.4 | 8 750 | 10.5 | -130 | [124] |
| | $CaAlSi_2O_8$ + 5 wt% $TiO_2$ | 950, 0.5h | 8.0 | 22 500 | 10 | -20 | [207] |
| | $NaAlSi_3O_8$ | 1025, 15h | 6.3 | 8 700 | 10.5 | -24 | [124] |
| | $Na_{0.8}Ca_{0.2}Al_{1.2}Si_{2.8}O_8$ | 1100, 15h | 5.8 | 17 600 | 10.5 | -7 | [124] |
| | $BaAl_2Si_2O_8$ | 1500, 40h | 7.2 | 77 000 | 11 | -22 | [125] |
| | $BaAl_2Si_2O_8$ | 1475, 3h | 6.36 | 44 800 | | -46.9 | [208] |
| | $BaZnSi_3O_8$ | 1100, 3h | 6.6 | 52 401 | 15.4 | -24.5 | [209] |
| | $SrZnSi_3O_8$ | 1150, 3h | 6.12 | 78 064 | 12-14 | -33.2 | [210] |
| | $Ba_{0.95}Ca_{0.05}Al_2Si_2O_8$ | 1425, 5 | 6.65 | 24335 | 8 | -19.3 | [211] |

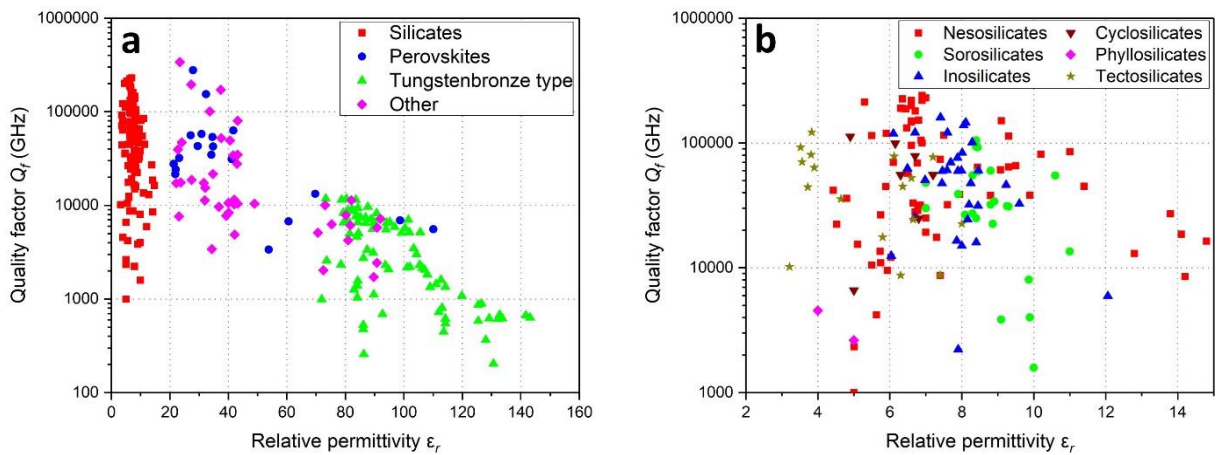

*Figure 6: a) Scatter plot of combinations of quality factor and permittivity showing silicates compared with other ceramic materials, based on a review of microwave ceramics [53]. b) Property combinations for silicates, further distinguished by silicate families.*

**Synthesis**

Silicate materials can be formed from oxide salts or organometallic precursors. The selection of processees and materials is directed by considerations of achieving dense defect-free microstructures without cost-prohibitive reagents and processes.

For the determination of dielectric properties as tabulated here, methods most commonly involve the use of dense disc-shaped ceramic samples produced via standard solid-state processing methods. This route involves the mixing of precursor powders, usually metal oxides, and their solid-state pre-reaction. The product is generally re-ground, pressed into a pellet, and sintered into a dense puck or pellet. For the milling steps, deionized water should be the preferred dispersing medium, as milling with ethanol might compromise the densification when working with carbonate precursors [208].

Although many studies have been dedicated to the topic of high frequency dielectrics, it is surprising that very little attention has been paid to the role of ceramic synthesis methods in governing dielectric performance. In fact, only 9 out of the 96 publications surveyed in Table 3 employed a synthesis or fabrication technique differing substantially from the standard method described above. Methods have been reported whereby melted oxide precursor powders were milled following quenching, with the resulting glass powdered being subsequently sintered [119,131,197,200]. This energy-intensive technique is found to yield improved loss characteristics in some cases, but may lead to compromised densification behaviour. Tsai et al. demonstrated that the pre-reaction/calcination step in the standard procedure can be omitted by employing a direct reaction-sintering approach for $Zn_{1.95}Co_{0.05}SiO_4$ ceramics while still obtaining very high $Q_f$ values [155]. This result was an outlier, though, as all other compounds performances in this study were incomparably worse. In other work, improved $\tau_f$ characteristics were found in liquid phase deposited $MgSiO_4$ layers [141]. However, this improvement can be attributed to the addition of $TiO_2$ rather than the production method itself.

The most promising modification to standard solid-state processing routes seems to be the synthesis of the precursor materials. This has been demonstrated by several studies [109,110,130], in which sol-gel synthesis methods were used to produce precursor powders instead of using commercially available metal oxide powders. The resulting benefits of these attempts include ultra-high quality factors, which are attributed to finer and phase purer starting powders, and in one case a reduced sintering temperature [128]. The reduction in sintering temperature is of further value in the fabrication of mmW circuits incorporating LTCCs.

The results from alternative synthesis approaches clearly indicate that many possibilities for improvement in terms of microstructure and dielectric properties in silicates remain unexplored. In particular, rather than using raw materials in the form of binary compounds (metal oxides mostly) in solid-state processing, the chemical synthesis of homogenous and pure ceramic powders of silicate ceramics followed by subsequent mechanical and thermal processing presents alluring avenues towards the attainment of high levels of dielectric performance. Although various sol-gel methods have been applied, the multitude of available soft chemistry powder synthesis routes and their effects on the relevant characteristics have not yet been investigated by researchers for silicate dielectrics. Besides this, the employment of more elaborate sintering techniques [212] than the standard pressing and furnace-firing could contribute to an advancement in the field.

## 7. Discussion

Figure 7 shows reported silicate dielectric property combinations, as tabulated in Table 3. Figures a, c and e differentiate materials according to their silicate family, whereas Figures b, d, f show the behaviour of single phase silicates vs. mixed phase materials. As can be expected, identifying clear trends from the broad range of studies collated here is challenging, however, a few assessments can be made:

- The highest quality factors have been achieved for Nesosilicates (up to 250.000 GHz for Forsterite, $Mg_2SiO_4$ [132]), while, with the exception of indialite [198], no ultra-high values (>> 100.000 GHz) have been reported for either Soro-, Cyclo-, Phyllo- or Tectosilicates. Almost all Nesosilicates are of lower or the same symmetry (orthorhombic) as compounds of the silicate groups named above, e.g. Hardystonite ($Ca_2ZnSi_2O_7$, tetragonal), Cordierite ($Mg_2Al_4Si_5O_{18}$, orthorhombic) or Quartz/Cristobalite/Tridymite ($SiO_2$, trigonal/tetragonal/orthorhombic), which reinforces the general assessment that for a given material, extrinsic loss factors tend to dominate intrinsic ones (symmetry). It should be considered here though, that the number of surveyed studies of Neso- and Inosilicates outweighs that for other silicate families.
- The lowest relative permittivity values (<4) have been reported for $SiO_2$ materials. This is not surprising since these materials consist purely of a $SiO_4$-tetrahedra network, resulting in extremely low dielectric polarizability values. However, no ultra-high $Q_f$ have been achieved for polycrystalline $SiO_2$ ceramics and because the structure contains no metal interstitials, cation-substitutions are limited to Si, rendering its tunability insignificant compared to other silicates.
- Although dielectric losses for phase pure ceramics are generally lower, it is still possible to obtain ultra-high $Q_f$ values for multi-phase materials. Noteworthy here are the compounds $Mg_2SiO_4$ + 1 wt% $TiO_2$ with $Q_f$ = 230.000 GHz which is almost as high as without $TiO_2$ (240.000 GHz) [139] and $LiAlSiO_4$ + 12 mol% $B_2O_3$ with $Q_f$ = 212.000 GHz [158].
- Similarly, although in most cases near zero $\tau_f$ for silicates ceramics are realized by the addition of a material with very high $\tau_f$, researchers demonstrated the possibility to design single phase materials with $\tau_f \approx 0$. Exemplary is the material $Mg_{1.92}Cu_{0.08}SiO_4$ with $\tau_f$ = -2 [144].
- Even though clear delimitations can not be made, a general trend in terms of relative permittivity can be observed from Figures 7a and 7e as follows: $\varepsilon_{r\ Tecto} < \varepsilon_{r\ Neso} < \varepsilon_{r\ Ino} < \varepsilon_{r\ Soro}$
The limited amount of data points for Cyclo- and Phyllosilicates does not allow an assessment of the comparative dielectric performance of these silicate families.

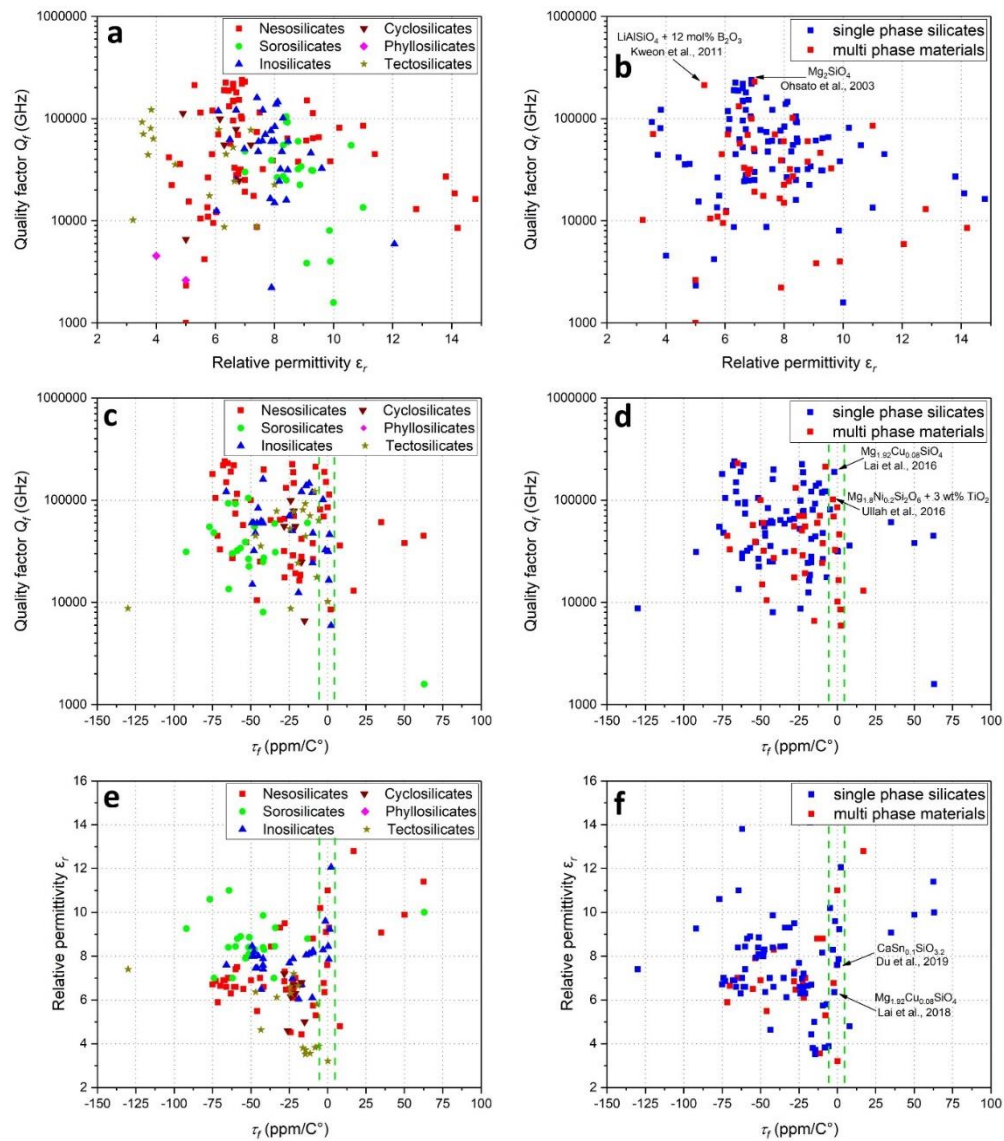

*Figure 7: Dielectric property combinations for silicate ceramics surveyed from the literature. The graphs in the left column divide the materials based on their silicate supergroup, while graphs in the right column differentiate in single- or multiphase ceramics.*

Although some general trends can be identified, certain ambiguity remains, including optimal sintering procedures and structure-property relations. However, for the production of high-performance silicate dielectrics ceramics, some guidelines can be formulated:

- In applications where geometries and operating frequencies of dielectric components are invariable, a compound with the desired permittivity can be designed by cation substitution using approaches described by Roberts [43] or Phillips, Van Vechten and Levine [50] described in section 3.

- Starting materials should be of high purity to prevent the formation of undesired secondary phases. In the case of solid-state reaction pathways, educt materials with high reactivity should be chosen, e.g., amorphous instead of crystalline $SiO_2$ or $Mg(OH)_2$ instead of $MgO$.

- If not deliberately targeted, porosity should be kept as low as possible to avoid deterioration of $Q_f$ values and deviation of $\varepsilon_r$ from expected levels. Measures to achieve optimal densification include

- precompaction via pressing to increase reactivity and appropriate sintering temperature (as low as possible), duration and cooling.
- The formation of liquid phases during sintering should be averted, as possible glassy phases in the final product are known to be detrimental to high $Q_f$ and, also, lead to unintended relative permittivities.
- Generally, a product with larger grains should be preferred as reduced interface polarization yields lower dielectric losses. However, abnormal grain growth will in turn also lead to compromised $Q_f$ values. Optimal sintering conditions depend on a multitude of factors, including materials parameters and the intended application, including the interaction with mm waves.
- Optimizing reaction and sintering conditions should have priority over trying to microtune material intrinsic loss mechanisms, as these only play a subordinate role in the total dielectric loss of a material.

Multiphase materials, although generally assumed to have greater losses, show some promise here owing to their lower processing temperatures, and the relative ease with which $\tau_f$ values close to zero can be achieved. The fabrication of integrated devices through co-firing of various ceramics with connectors using LTCCs, necessitates the use of compositions that can be densified without prohibitively high temperatures. This remains a problematic aspect of silicate dielectrics, as most of the materials surveyed here involve the use of temperatures over 1100°C, and as high as 1500°C that are too high to be compatible with other LTCCs and metallic connectors. It is clear that a reduction of thermal processing temperatures can be realised in silicate ceramics by employing soft chemistry-based synthesis methods rather than standard solid-state processing. Further exploring these routes yields dividends on multiple fronts and remains a high value objective in ceramic research. The implementation of soft-chemistry methods may in the future allow for mmW co-firable silicate dielectrics, which are currently not commercially available.

In order to minimise extrinsic dielectric losses in silicate ceramics, a combination of high density and reduced interface polarisation at grain boundaries should be targeted. As with the engineering of permittivity and the reduction of thermal processing temperatures, soft-chemistry methods have a role to play here too. Local chemistry at grain boundaries, phase segregation behaviour, and oxygen vacancies all significantly impact the extrinsic component of dielectric losses. Minimising these losses involves tailoring chemical and thermal processing to minimise interface polarisation at grain boundaries and maximise local symmetry within the lattice of a single phase. The distribution of substitutional cations can further be studied by DFT methods to identify energetically favourable configurations and local structure's role on the dielectric behaviour.

With the accelerating implementation of large numbers of mmW components in diverse applications utilising ever increasing frequencies in the electromagnetic spectrum, dielectric materials' requirements are undergoing some shifts. The role of new low permittivity ceramic dielectric materials is still minor in existing applications, however, the current trend suggests that such materials are likely to experience increasing demand in terms of volume and performance. To address such demand, silicate ceramics present some attractive avenues forwards. Specifically, the design of new silicate materials and methods for their processing is a worthwhile undertaking with the goal of obtaining novel low-loss dielectric with well-tuned and thermally stable permittivity. To engineer new silicate materials, better understandings are required to establish the role of lattice structure and cation occupancy on ionic and electronic polarization mechanisms that govern the response of a material to an electric field in the mmW region of the spectrum, in the form of the material's permittivity and quality factor. The ability to substitute a broad range of cations at multiple sites in the different silicate families, coupled with the compound effects wrought by such substitutions on the high frequency polarization and intrinsic dielectric losses as discussed above, motivate further computational and experimental investigations into substituted silicate ceramics in mmW dielectric applications.

## 8. Conclusion

As an approach to address the rapidly growing need for well-controlled low-*k* dielectrics in emerging millimetre wave applications, silicate ceramics present attractive attributes. Consequently, they have emerged as the focal point for numerous studies towards such applications, which have been reviewed here. The surveyed materials' attributes in terms of relative permittivity, dielectric loss, and temperature factor of resonant frequency show that silicates are inexpensive material candidates for applications in the top range of microwave frequencies as dielectric resonators, waveguides, filters, and substrate materials. The most immediate barrier towards the utilisation of silicate dielectric ceramics is that this class of materials is currently limited in terms of reported compositions combining appropriate dielectric performance values with sintering temperatures <1000°C, necessary for the implementation in LTCCs. With a view towards advancing this new class of materials, we identify here several key objectives for future research in the field:

- The exploration of soft chemistry routes for the production of phase pure and homogeneous ceramic powders and their effect on product ceramics in terms of microstructure and dielectric properties
- The development of low temperature co-firable low-loss silicate dielectrics
- The revaluation of crystal chemical loss factors, e.g. how the degree of $SiO_4$-tetrahedra polymerisation governs dielectric losses in silicates.

**Funding**

This research did not receive any specific grant from funding agencies in the public, commercial, or not-for-profit sectors.